\documentclass[preprint,showpacs,preprintnumbers,amsmath,amssymb]{revtex4}
\usepackage{graphicx}

\begin{document}


\title{Arbitrary-order non-linear contribution to self-steepening}

\author{J. Kasparian$^{1,*}$, P. B\'ejot$^2$, J.-P. Wolf$^1$}
\affiliation{(1) Universit\'e de Gen\`eve, GAP-Biophotonics, 20 rue
de l'Ecole de M\'edecine, 1211 Geneva 4, Switzerland}
\affiliation{(2) Laboratoire Interdisciplinaire Carnot de Bourgogne (ICB), UMR 5209 CNRS-Universit\'e de Bourgogne, 9 Av. A. Savary, BP 47 870, F-21078 DIJON Cedex, FRANCE}
\affiliation{$^*$ Corresponding author: jerome.kasparian@unige.ch}


\begin{abstract}
Based on the recently published generalized Miller formul\ae, we derive the spectral dependence of the contribution of arbitrary-order non-linear indices to the group-velocity index. We show that in the context of laser filamentation in gases all experimentally-accessible orders (up to the $9^{th}$-order non-linear susceptibility $\chi^{(9)}$ in air and $\chi^{(11)}$ in argon) have contributions of alternative signs and similar magnitudes. Moreover, we show both analytically and numerically that the dispersion term of the non-linear indices must be considered when computing the intensity-dependent group velocity.
\end{abstract}

 
\maketitle


Non-linear optics \cite{Boyd} relies on the non-linear properties of the propagation medium, among which the successive orders of the non-linear susceptibility are essential parameters. However, due to the difficulty to measure them experimentally, their knowledge is generally limited to the first non-zero order ($\chi^{(2)}$ or $\chi^{(3)}$, depending on medium symmetry). Furthermore, the available laser sources drastically limit the wavelengths available for such measurements, so that reliable dispersion curves for higher-order susceptibilities cannot be deduced from the sparse experimental data available to date. 

The lack of data led to neglect these higher-order Kerr terms in most numerical simulations of e.g. both self-guided filaments in ultrashort intense laser pulses \cite{KasparianW08,BergeSNKW07,CouaironM07,ChinHLLTABKKS05} or the propagation of high-intensity pulses in hollow-core fibers \cite{DudleyGC06}. Similarly, the Kerr contribution to the group velocity is most generally limited to the third-order and treated as dispersionless in the lack of data about its dispersion \cite{BergeSNKW07,CouaironM07}.
Recently, however the measurement of the higher-order refractive indices up to $n_{8}$ in N$_2$ and O$_2$, and up to $n_{10}$ in argon \cite{LoriotHFL09}, followed by the generalization of the Miller formul\ae\ \cite{Miller64} to any order of non-linearity \cite{EttoumiPKW10}, provided a new insight into the spectral dependence of the non-linear refractive index at high incident intensity.

This allowed us to show that these terms cannot be neglected and can even provide the dominant contribution stabilizing the self-guided filaments \cite{LoriotBHFLHKW09}. Furthermore, in argon-filled hollow-core fibers, these terms are necessary to obtain quantitative agreement of numerical simulations with experimental data \cite{SchmidtBGSTBKWVKCL10,BejotSKWL10}. But these works focused on the contribution of higher-order Kerr terms to phase velocity. The contribution of the spectral dispersion of higher-order indices to group velocity was not considered, although it can be expected to impact the propagation, and in particular the self-steepening term.


Here, we derive an explicit expression for the contribution of any order of the non-linear refractive indices to the group-velocity index. We show that in filamentation, all non-linear orders of the group-velocity index can have similar orders of magnitude and must be considered. Furthermore, especially in the ultraviolet, the dispersion of the non-linear indices cannot be neglected when calculating self-steepening.

At arbitrary intensity $I$ and frequency $\omega$, the refractive index can be expressed as \cite{Boyd}:
\begin{equation}
n(\omega)=n_0(\omega)+n_2(\omega)I+n_4(\omega)I^2+...=\sum_{j=0}^{\infty}{n_{2j}(\omega)I^j} \label{indice}
\end{equation}
In gases, where $n-1 \ll 1$, the refractive index $n(\omega)=\sqrt{1+\sum_{j=0}^{\infty}{\chi^{(2j+1)}(\omega)}I^j}$  can be 
approximated by \cite{Boyd}:
\begin{eqnarray}
n_0(\omega)&\approx&1+ \frac{1}{2}\chi^{(1)} (\omega) \\
n_{2j}(\omega)&\approx&Z^{(2j+1)}\chi^{(2j+1)}(\omega)
\label{DefZ}
\end{eqnarray}
where $\chi^{(2j+1)}$ is the $(2j+1)^{th}$-order non-linear susceptibility and the $Z^{(2j+1)}$ are frequency-independent factors.
If a pulse can be described as a carrier wave modulated by an envelope with a sufficiently narrow spectrum to allow neglection of the envelope deformations over short distances, then a group-velocity index can be defined as
\begin{equation}
 n_g(\omega)=\frac{c}{v_g}=n(\omega)+\omega \frac{\textrm{d}n}{\textrm{d}\omega}  \label{definition_n_g} 
 \end{equation}
where $v_g$ is the group-velocity and $c$ is the speed of light in vacuum. Defining $Z^{(1)}=1/2$ and considering Equations (\ref{indice})-(\ref{DefZ}), $n_g$ rewrites:

\begin{eqnarray}
n_g(\omega)&=&\sum_{j=0}^{\infty}{\left(n_{2j}(\omega)+\omega\frac{dn_{2j}}{d\omega}\right)I^j}\equiv \sum_{j=0}^{\infty}{n_{g,2j}I^j} \\
&\approx&1+\sum_{j=0}^{\infty}{Z^{(2j+1)}\left(\chi^{(2j+1)}(\omega)+\omega\frac{d\chi^{(2j+1)}}{d\omega}\right)I^j}
\label{def_n_g}
\end{eqnarray}

Identifying the terms for each power of the intensity, we obtain the contribution of each order of non-linearity to the group-velocity index.

\begin{eqnarray}
n_{g,0}(\omega) -1 &\approx& \frac{1}{2}\left(\chi^{(1)}+\omega\frac{d\chi^{(1)}(\omega)}{d\omega}\right)
\label{ng_chi1} \\
\forall j\ge 1, n_{g,2j}(\omega) &\approx& Z^{(2j+1)}\left(\chi^{(2j+1)}+\omega\frac{d\chi^{(2j+1)}(\omega)}{d\omega}\right)
\label{ng_chi_q} 
\end{eqnarray}

The first expression corresponds to the usual linear contribution, while the first non-linear contribution to the group-velocity index $n_{2,g}$ corresponds to the classical self-steepening term \cite{Boyd,AkozbekSBC01,JOT02}. Within the elastically bound electron model, the susceptibility of abitrary order is given up to any order by generalized Miller formul\ae\ \cite{EttoumiPKW10}: 
\begin{eqnarray}
\chi^{(1)}(\omega) &=& \frac{N e^2}{m \epsilon_0({\omega_0}^2-\omega^2+i\omega_0\gamma)}  \label{Sellmeier1} \\
\forall q \ge 2,\ \ \chi^{(q)}(\omega) 
&=&\frac{Ne}{\epsilon_0}\left(\frac{e}{m}\right)^q Q^{(q)} \frac{1}{\Omega(\omega)^{q+1}}
\label{chi_q}
\end{eqnarray}
where $m$ and $-e$ are the electron mass and charge, $\epsilon_0$ the permittivity of vacuum, $N$ the density of dipoles in the propagation medium, $\gamma$ the width of the resonance at frequency $\omega_0$ and $Q^{(q)}$ describes the potential well where the electron oscillates; $\Omega(\omega)=\omega_0^2-\omega^2+i\omega_0\gamma$.
Inserting these expressions into Equations (\ref{ng_chi1}) and (\ref{ng_chi_q}) yields:
\begin{eqnarray}
n_{g,0}(\omega)-1
&\approx&\left(n_0(\omega)-1\right)\frac{\omega_0^2+\omega^2+i\omega_0 \gamma}{\omega_0^2-\omega^2+i\omega_0 \gamma}
\label{n_g0} \\
\forall j \ge 1,
n_{g,2j}(\omega)
&\approx& n_{2j}(\omega) \frac{\omega_0^2+(4j+3)\omega^2+i\omega_0 \gamma}{\Omega(\omega)}
\label{ng_2j}
\end{eqnarray}

This equation provides a general expression of each non-linear contributions to the group-velocity index $n_g$. As a consequence, it allows to evaluate the impact of higher-order Kerr terms on the self-steepening in the context of laser filamentation in gases or the propagation of ultrashort pulses in hollow fibers. In the following, we consider the propagation of high-intensity pulses in transparent media, far from resonance. In this case, $|\omega_0-\omega| \gg \gamma$, so that $\omega_0^2+\omega^2 > |\omega_0^2-\omega^2| \gg \omega_0\gamma$. The imaginary parts of Equations (\ref{n_g0}) and (\ref{ng_2j}) become negligible:
\begin{eqnarray}
n_{g,0}(\omega)-1&\approx& (n_0(\omega)-1)\frac{\omega_0^2+\omega^2}{\omega_0^2-\omega^2} \label{n_g0final} \\
n_{g,2j}(\omega)&\approx& n_{2j} \frac{\omega_0^2+(4j+3)\omega^2}{\omega_0^2-\omega^2} \label{n_g2final}
\end{eqnarray}

\begin{figure}[t]
  \begin{center}
      \includegraphics[keepaspectratio, width=8cm]{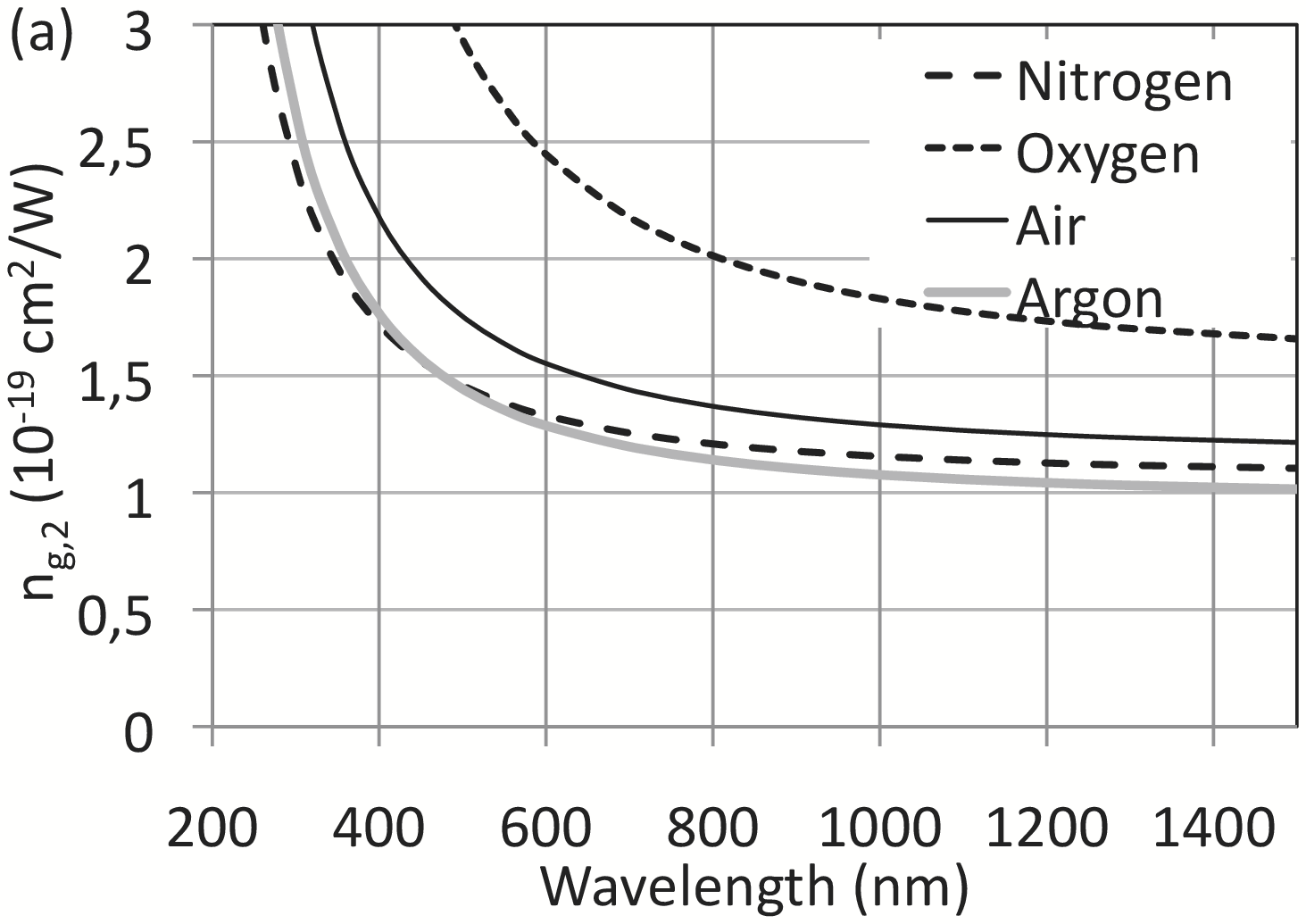}
      \includegraphics[keepaspectratio, width=8cm]{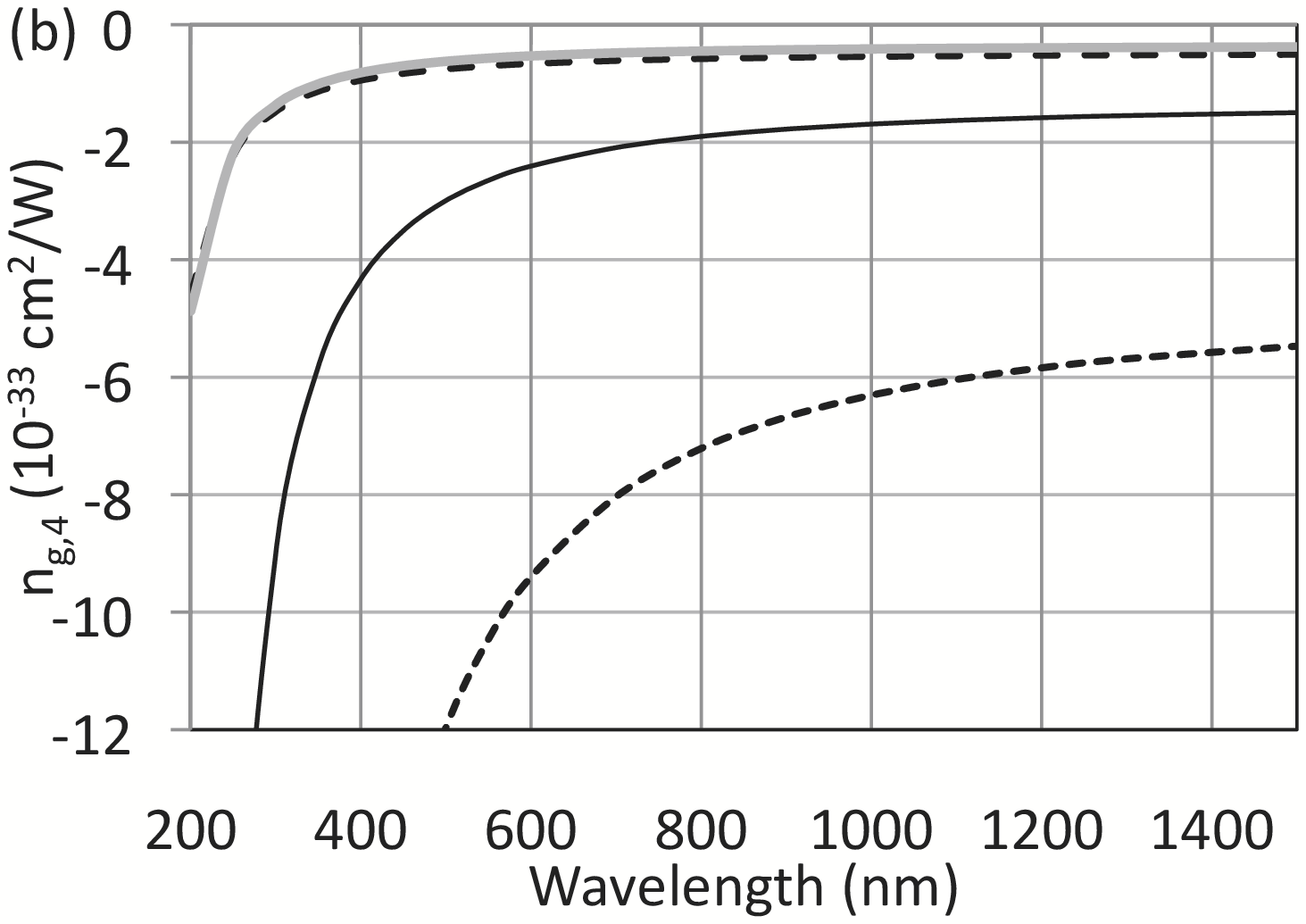}
      \includegraphics[keepaspectratio, width=8cm]{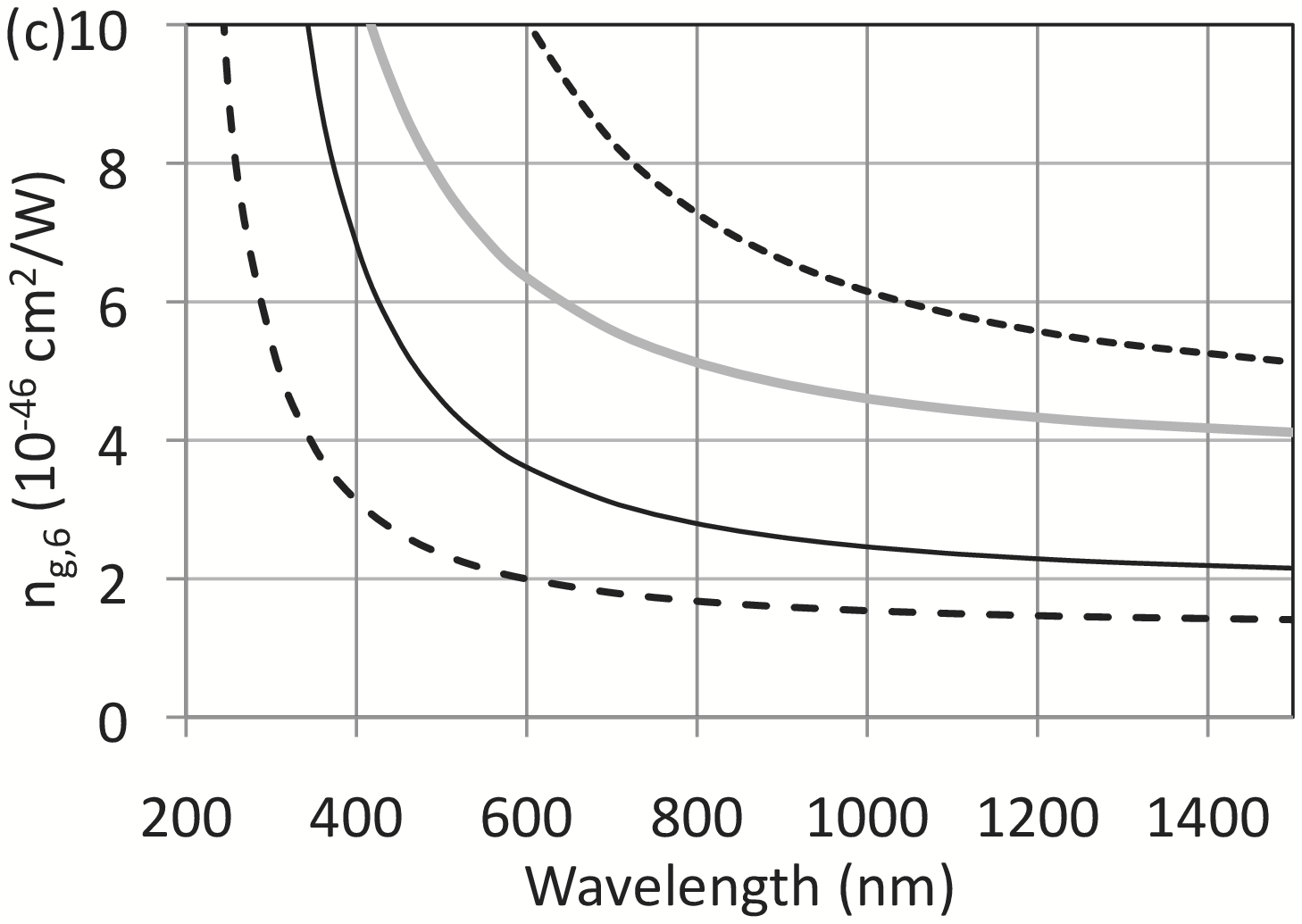}
      \includegraphics[keepaspectratio, width=8cm]{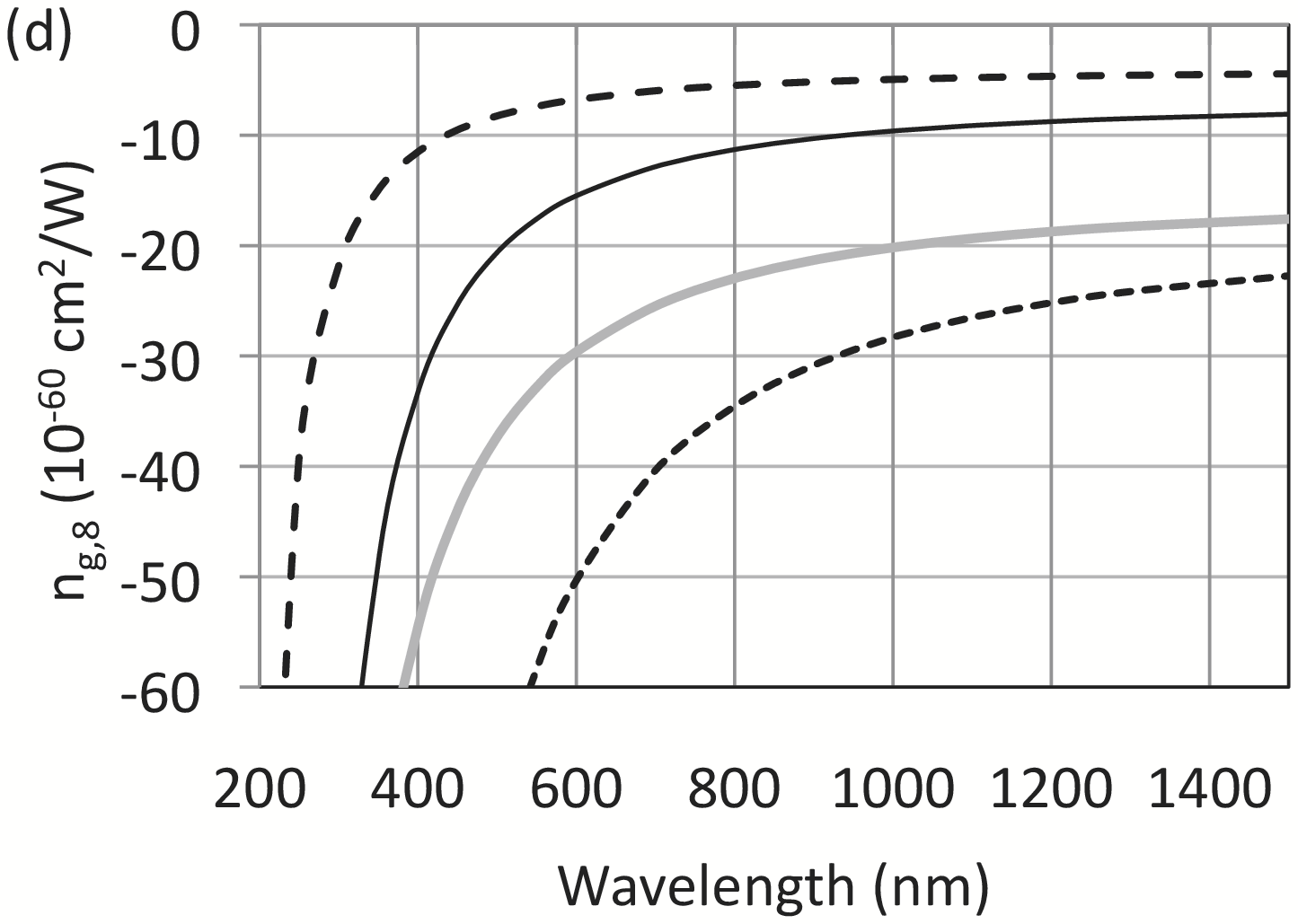}
  \end{center}
  \caption{Spectral dependence of the non-linear group-velocity indices (a) $n_{g,2}$, (b) $n_{g,4}$, (c) $n_{g,6}$, and (d) $n_{g,8}$ of O$_2$, N$_2$, air and Ar at atmospheric pressure}
  \label{n_gi}
\end{figure}

Note that the negative values obtained for $\omega > \omega_0$ correspond to the well-known region of negative group-velocity index \cite{BoydG09}. Figure \ref{n_gi} displays the spectral dependence from Equation (\ref{n_g2final}), based on the recent experimental measurements of $n_{2j}$ at 800 nm \cite{LoriotHFL09,LoriotBHFLHKW09}, extrapolated to the whole visible spectrum by applying generalized Miller formul\ae\ \cite{EttoumiPKW10} and the dispersion data of Zhang \emph{et al.} \cite{ZhangLW08}. 

From Equations (\ref{n_g0final}) and (\ref{n_g2final}), we can estimate the ratio of the successive terms of the group-velocity index 
\begin{eqnarray}
\frac{n_{g,2}I}{n_{g,0}-1} &=& \frac{n_{2}I}{n_0-1} \ \frac{\omega_0^2+7\omega^2}{\omega_0^2+\omega^2} \label{ratio_0}\\
\forall j \ge 1, \frac{n_{g,2j+2}I^{j+1}}{n_{g,2j}I^j} &=& \frac{n_{2j+2}I}{n_{2j}}\ \frac{\omega_0^2+(4j+7)\omega^2}{\omega_0^2+(4j+3)\omega^2} \label{ratio_j}
\end{eqnarray}

\begin{figure}[t]
  \begin{center}
      \includegraphics[keepaspectratio, width=8cm]{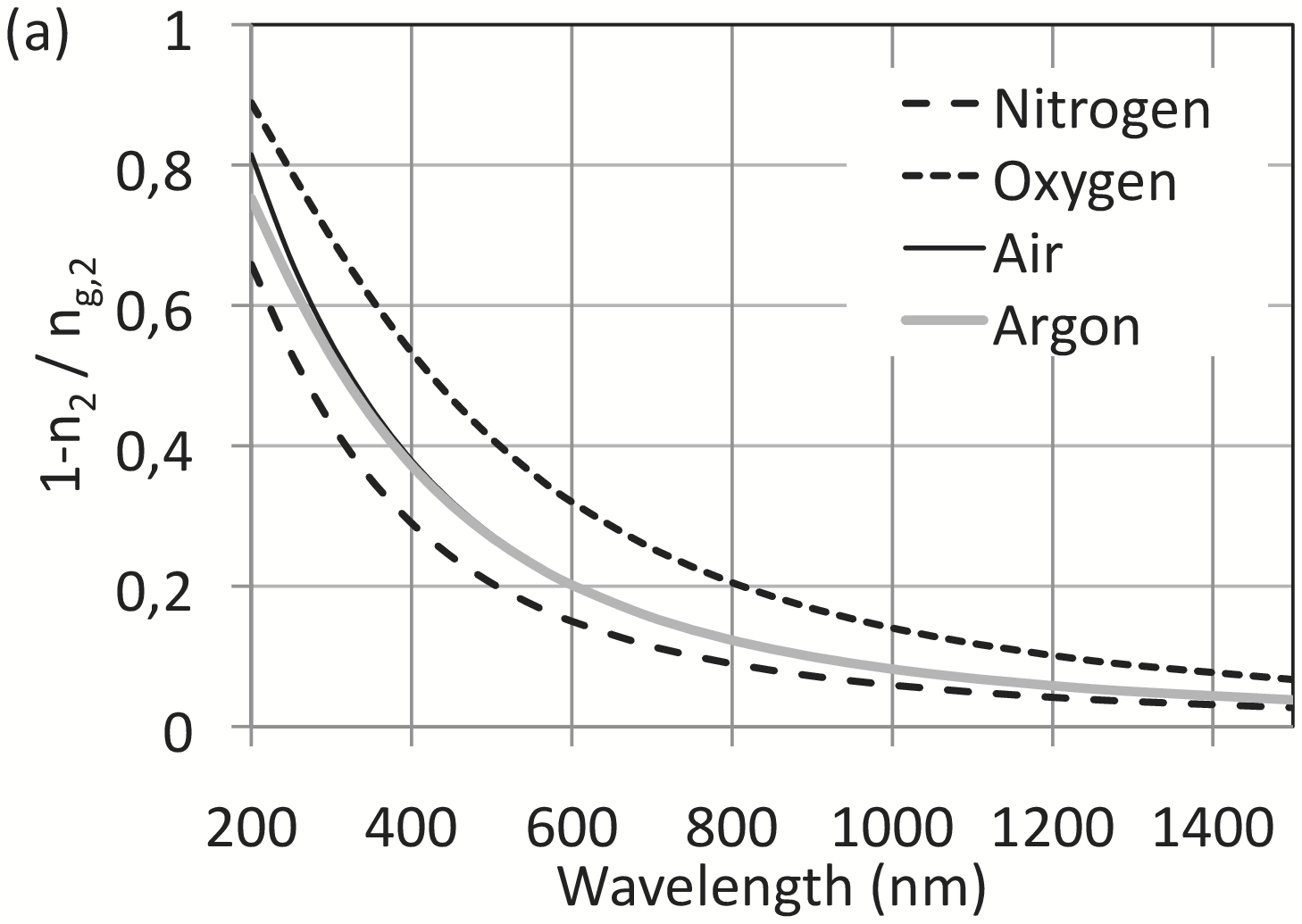}
      \includegraphics[keepaspectratio, width=8cm]{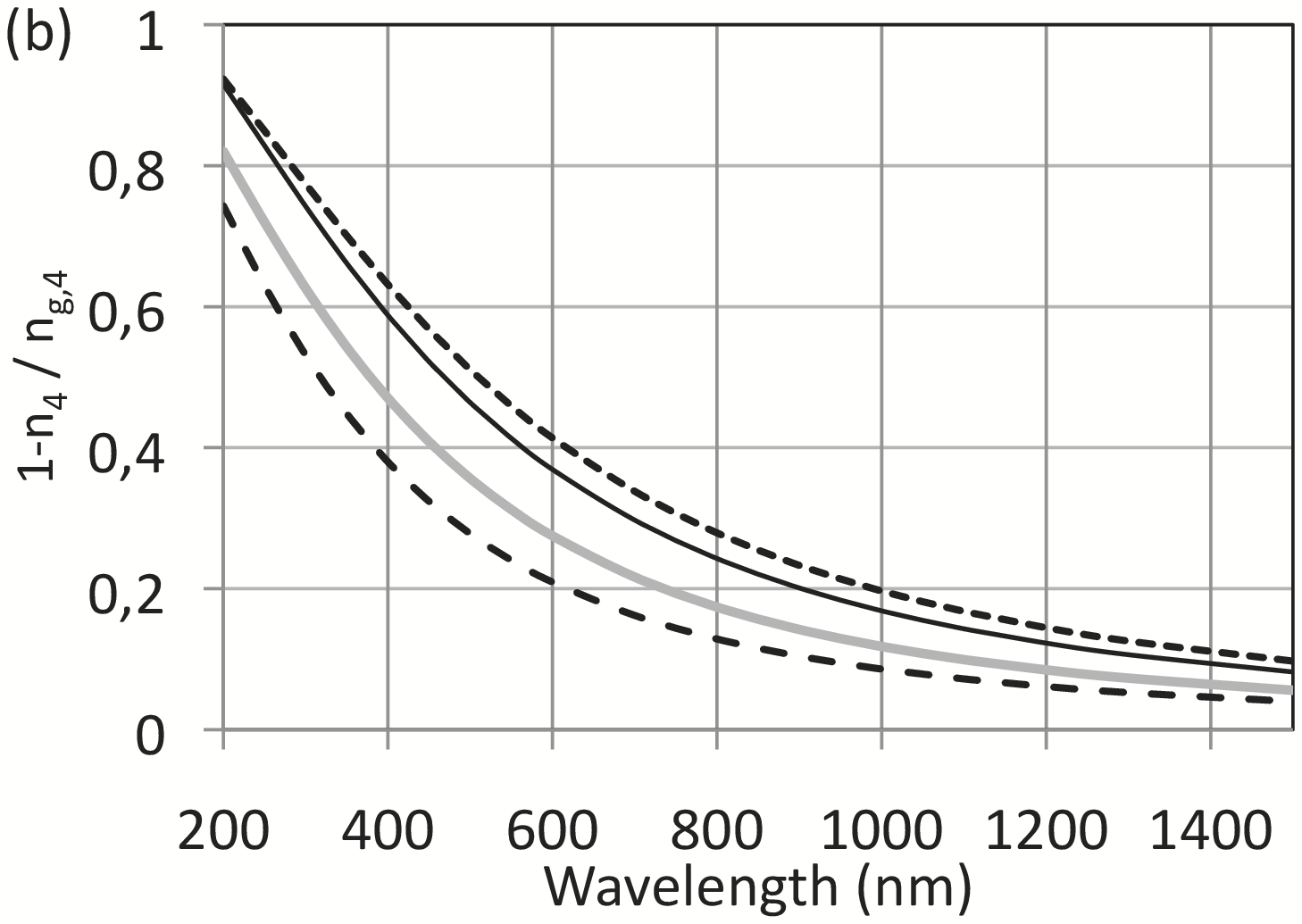}
      \includegraphics[keepaspectratio, width=8cm]{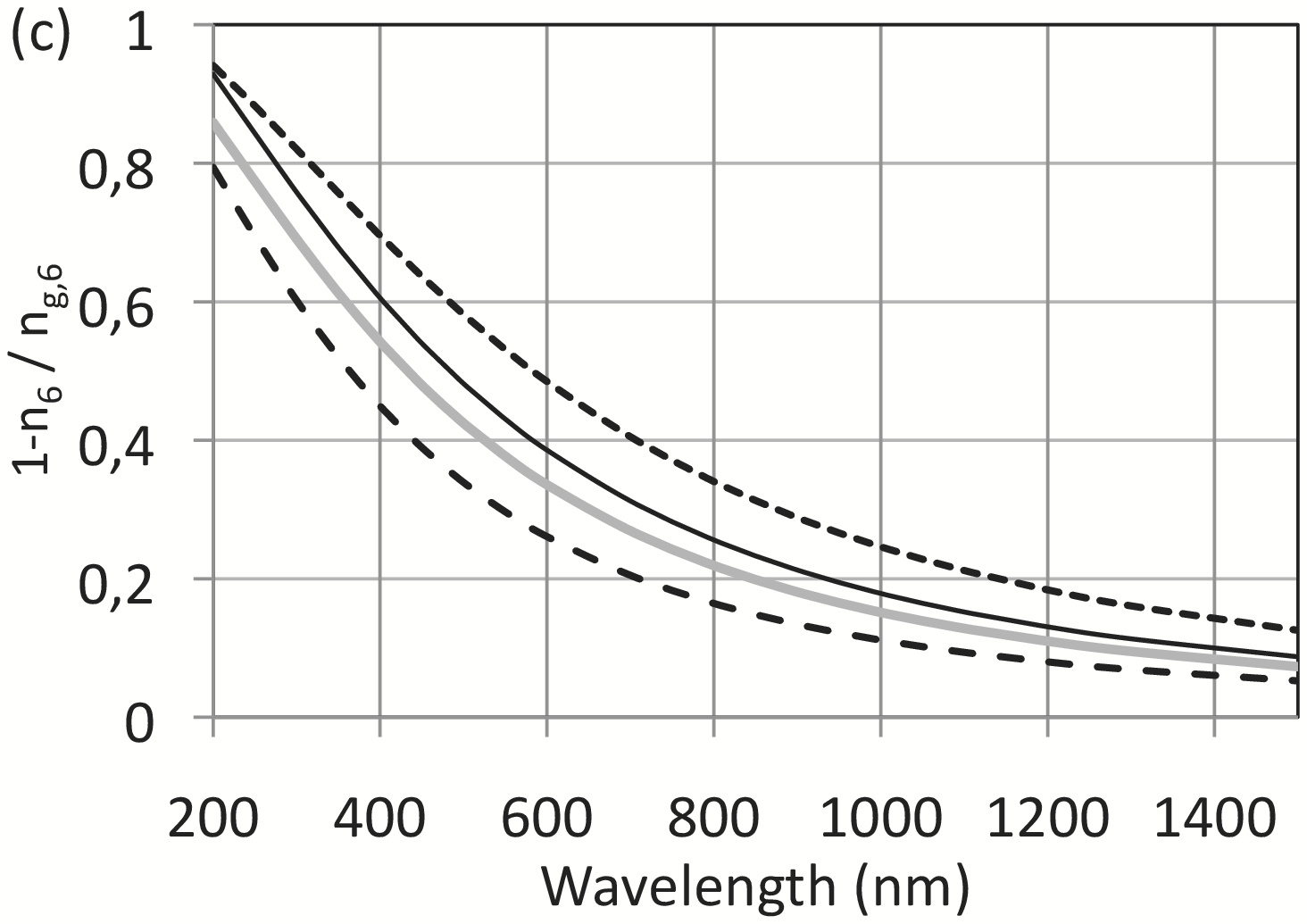}
     \includegraphics[keepaspectratio, width=8cm]{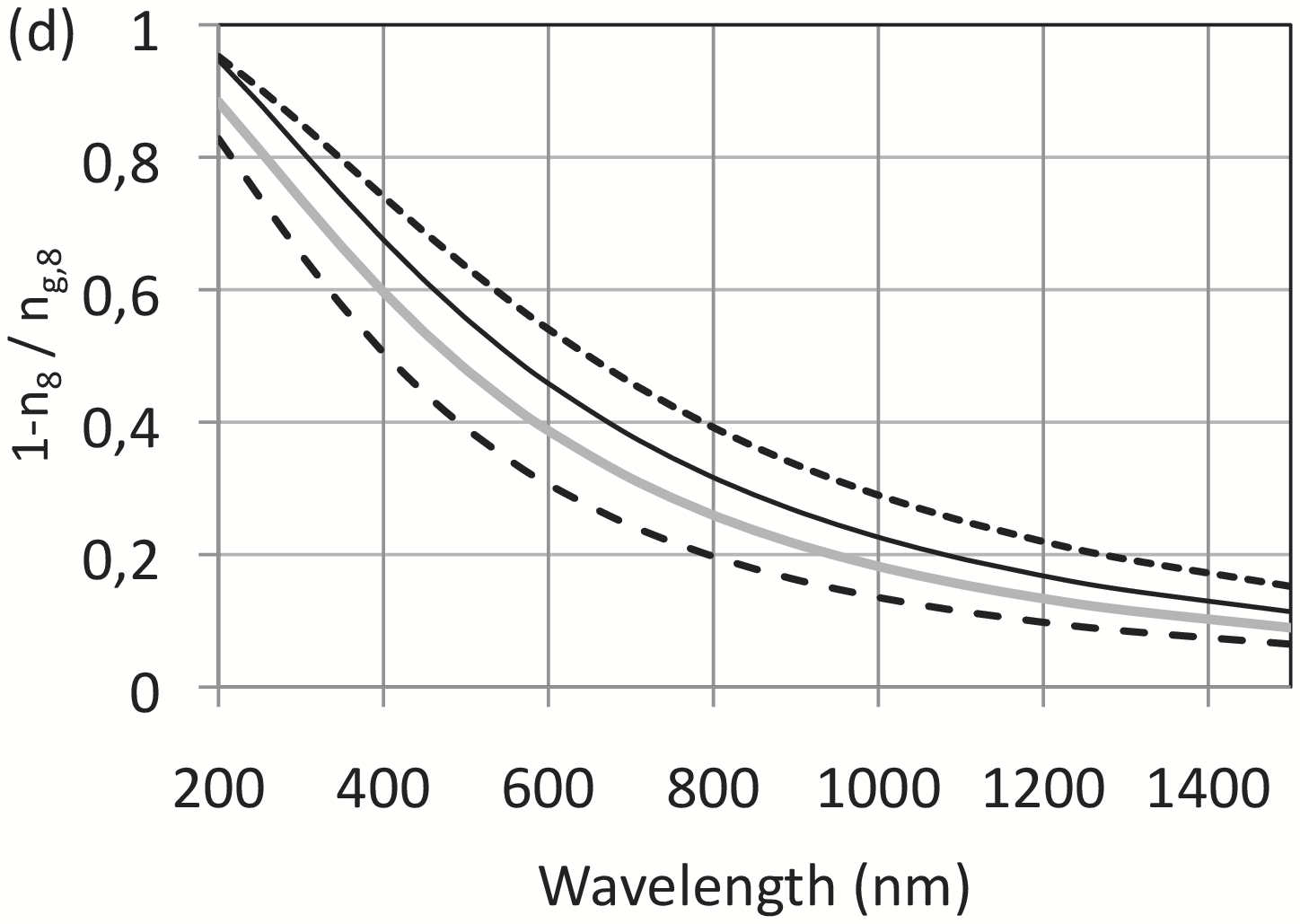}
  \end{center}
  \caption{Relative error induced when neglecting the dispersion of the Kerr terms in the group-velocity (a) $1-n_2/n_{g,2}$, (b) $1-n_4/n_{g,4}$, (c) $1-n_6/n_{g,6}$, and (d) $1-n_8/n_{g,8}$ 
of O$_2$, N$_2$, air and Ar at 1 atm.}
  \label{erreur_relative}
\end{figure}

The second factor of equations (\ref{ratio_0}) and (\ref{ratio_j}) is of the order of 1. Therefore, the orders of magnitude of the ratio of successive terms are driven by the ratio of the non-linear indices $n_{2j}$, multiplied by $I$. The values displayed in Figure \ref{n_gi} imply that, for $I < 10^{14}$ W/cm$^2$, $n_{2j}I^j \ll n_0-1$. 
Self-steepening is therefore, as well-known \cite{AkozbekSBC01,JOT02}, a second-order term in the non-linear Schr\"odinger equation (NLSE) describing the non-linear propagation of light in a non-linear transparent medium. Furthermore, all known terms in $n_{2j}I^j$ have alternate signs and comparable orders of magnitude \cite{LoriotHFL09, EttoumiPKW10}. The same therefore applies to the terms in $n_{g,2j}I^j$, which must all be taken into account when describing self-steepening e.g. in the context of filamentation, where the intensity is clamped around $5\times10^{13}$ W/cm$^2$ \cite{KasparianSC00, BeckerAVOBC01}, or of the propagation in hollow fibers.


\begin{figure}[t]
  \begin{center}
      \includegraphics[keepaspectratio, width=8.5cm]{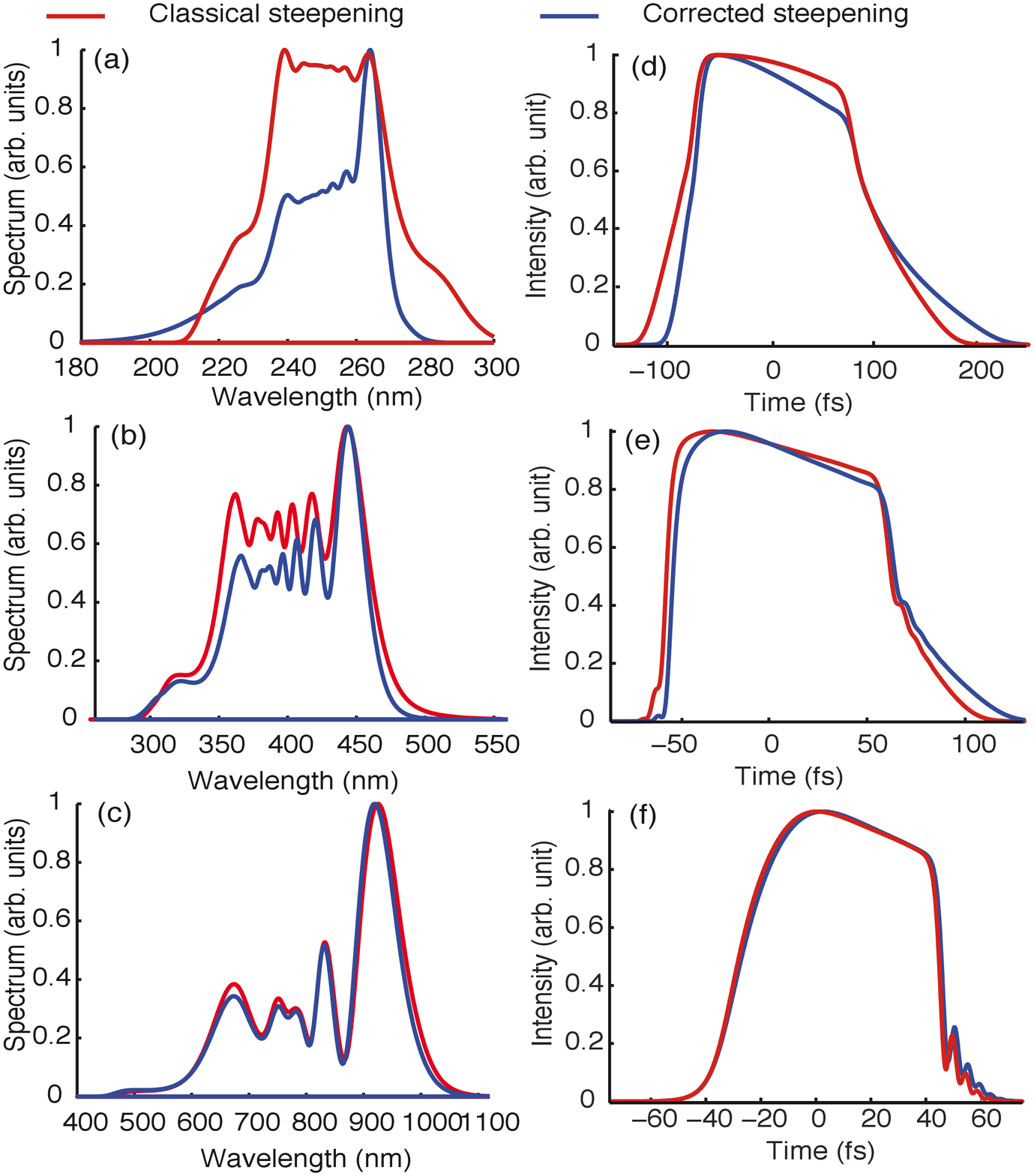}
  \end{center}
  \caption{Influence of the higher-order indices contribution to the group-velocity index on the propagation of a 35~fs, pulse with fixed $n_2(\lambda)I$ in a 1~m long hollow-core fiber filled with 1.4 bar argon. (a,d) 250 nm, 286 $\mu$J; (b,e) 400 nm, 360 $\mu$J; (c,f) 800 nm, 400 $\mu$J}
  \label{comparaison_spectre}
\end{figure}

Equation (\ref{n_g2final}) provides an estimation of the error performed when neglecting the dispersion term in the contribution of the higher-order indices to the group-velocity index. Figure \ref{erreur_relative} displays the relative error $1-n_{2j}/n_{g,2j}$ implied at atmospheric pressure when neglecting the dispersion terms of the Kerr contributions to the group velocity. The calculations are based on the same data as in Figure \ref{n_gi}. As is clear from Equation (\ref{n_g2final}), this error decreases for longer wavelengths where dispersion is smoother. At 800 nm, it amounts to  $\sim$20\% and may be considered acceptable, although not negligible. However, at blue or ultraviolet wavelengths, the dispersion term dominates and must be considered in the equations. 

Numerical simulations of the propagation of a 35 fs pulse in a  1 m long hollow-core fiber filled with 1.4 bar argon, confirm this finding. As described in detail earlier \cite{BejotSKWL10}, the model implements the NLSE including the higher-order Kerr terms. We compared the code output with and without the contribution of the higher-order indices to the group-velocity index up to the term in $n_{10}$, i.e. the terms of Equation (\ref{def_n_g}) for $1\le j \le 5$. As can be seen in Figure \ref{comparaison_spectre}, the consideration of the full steepening term affects the spectrum by deforming the pulse envelope. It simultaneously red-shifts the central part of the spectrum, and blue-shifts its edges. Furthermore, as predicted by the analytic calculations, the contribution of the dispersion of the higher-order Kerr terms is larger in the UV and negligible in the infrared. These these terms must therefore be considered in numerical simulations, especially while investigating spectral broadening.

In conclusion, based on the recent generalization of the Miller formul\ae, we have estimated the contribution of higher-order indices to the group-velocity index. These contributions define the self-steepening term of the NLSE. They have alternate signs and comparable absolute values in intensity regimes typical of filamentation. All non-linear terms must therefore be considered in the evaluation of the self-steepening of ultrashort intense laser pulses propagating in transparent Kerr media. Furthermore, we demonstrate both analytically and numerically that their spectral dispersion cannot be neglected either, especially at shorter wavelengths.

Acknowledgements. This work was supported by the Swiss NSF (contract 200021-125315) and the ANR COMOC project.
\bibliographystyle{unsrt}

\begin{thebibliography}{}

\bibitem{Boyd}
R. W. Boyd, \emph{Nonlinear optics}, Academic Press, 2008



\bibitem{ChinHLLTABKKS05}
S.~L. Chin, S.~A. Hosseini, W.~Liu, Q.~Luo, F.~Th\'eberge, N.~Ak\"ozbek, A.~Becker, V.~P. Kandidov, O.~G. Kosareva, H.~Schroeder.
\newblock {Can. J. Phys.} \textbf{83}, 863 (2005).

\bibitem{BergeSNKW07}
L.~Berg\'e, S.~Skupin, R.~Nuter, J.~Kasparian, J.-P.~Wolf.
\newblock {Rep. Prog. Phys.} \textbf{70}, 1633 (2007).

\bibitem{CouaironM07}
A.~Couairon A.~Mysyrowicz.
\newblock {Phys. Rep.}, \textbf{441} 47 (2007).

\bibitem{KasparianW08}
J.~Kasparian and J.-P. Wolf.
\newblock {Opt. Express} \textbf{16}, 466 (2008).

\bibitem{DudleyGC06}
J.M. Dudley, G. Genty, S. Coen, 
Rev. Mod. Phys. \textbf{78}, 1135 (2006)


\bibitem{LoriotHFL09}
V.~Loriot, E.~Hertz, O.~Faucher, B.~Lavorel,
Opt. Express \textbf{16}, 13429 (2009); Erratum in Opt. Express \textbf{18} 3011 (2010)

\bibitem{Miller64}
R. C. Miller,
Appl. Phys. Lett. \textbf{5}, 17 (1964)

\bibitem{EttoumiPKW10}
W. Ettoumi, Y. Petit, J. Kasparian, J.-P. Wolf, 
Opt. Express \textbf{18}, 6613 (2010)



\bibitem{LoriotBHFLHKW09}
P. B\'ejot,  J. Kasparian, S. Henin, V. Loriot, T. Vieillard, E. Hertz, O. Faucher, B. Lavorel, J.-P. Wolf,
Phys. Rev. Lett. \textbf{104}, 103903 (2010)

%

\bibitem{SchmidtBGSTBKWVKCL10}
B.E. Schmidt, P. B\'ejot, M. Gigu\`ere, A.D. Shiner, C. Trallero-Herrero, E. Bisson, J. Kasparian, J.-P. Wolf, D.M. Villeneuve, J.-C. Kieffer, P.B. Corkum, and F. L\'egar\'e, 
Appl. Phys. Lett. \textbf{96}, 121109 (2010)

\bibitem{BejotSKWL10}
P. B\'ejot, B. E. Schmidt, J. Kasparian, J.-P. Wolf, F. Legar\'e,
Phys. Rev. A \textbf{81}, 063828 (2010)

\bibitem{AkozbekSBC01}
N. Ak\"ozbek, M. Scalora, C.M. Bowden, and S.L. Chin, 
Opt. Commun. \textbf{191}, 353 (2001)

\bibitem{JOT02}
I. S. Golubtsov, O. G. Kosareva, 
J. Opt. Technol. \textbf{69}, 462 (2002)

\bibitem{BoydG09}
R. W. Boyd, D. J. Gauthier,
Science \textbf{326}, 1074 (2009)

\bibitem{ZhangLW08}
J. Zhang, Z. H. Lu, L. J. Wang,
Appl. Opt. \textbf{47}, 3143(2008)

%
%
%
\bibitem{KasparianSC00}
J.~Kasparian, R.~Sauerbrey, and S.~L. Chin.
\newblock {Appl. Phys. B} \textbf{71}, 877 (2000).

\bibitem{BeckerAVOBC01}
A.~Becker, N.~Ak\"ozbek, K.~Vijayalakshmi, E.~Oral, C.~M.~Bowden, S.~L. Chin.
\newblock {Appl. Phys. B} \textbf{73}, 287 (2001).










\end{thebibliography}

\end{document}